\documentstyle[12pt,aps]{revtex}
\title{Giant Coulomb broadening and Raman lasing \\ on ionic
transitions}

\author{A.A.~Apolonsky, S.A.~Babin, A.I.~Chernykh,
S.I.~Kablukov,\\ S.V.~Khorev, E.V.~Podivilov, D.A.~Shapiro}
\address{Institute of Automation \& Electrometry,\\
Novosibirsk, 630090, Russia}

\begin{document}
\draft
\maketitle
\begin{abstract}
CW generation of anti-Stokes Raman laser on a number of blue~-~green
argon-ion lines (4p--4s, 4p--3d) has been demonstrated with
optical pumping from metastable levels $3d'\,^2G, 3d\,^4F$.
It is found, that the population transfer rate is increased by
a factor of 3--5 (and hence, the output power of such Raman laser)
owing to Coulomb diffusion in the velocity space.  Measured are the
excitation and relaxation rates for the metastable level. The
 Bennett hole on the metastable level has been recorded using the
probe field technique. It has been shown that the Coulomb diffusion
changes shape of the contour to exponential cusp profile while its
width becomes 100~times the Lorentzian one and reaches values close
to the Doppler width. Such a giant broadening is also confirmed by
the shape of the absorption saturation curve.

\pacs{42.55.Ye, 52.20.Hv, 32.70.Jz}

\end{abstract}

\section{Introduction}
Stimulated Raman light scattering is widely used for frequency
conversion of laser radiation~\cite{WHITE}. Last years there is
 active development of CW anti-Stokes Raman lasers (ASRL) with
frequency up-conversion that use two-photon inversion in active
media of known lasers with gas-discharge excitation. CW generation
of the ASRL was successfully achieved for the media of argon-ion
\cite{WEL}, He-Ne \cite{WEL2} lasers and heavier noble atoms in
electric discharge~\cite{WEL3}. Conversion of radiation
(648.3~nm$\to$437.5~nm) in a 3-level $\Lambda$-scheme was studied
\cite{WEL} for ArII:  $3d\;^2P_{3/2}\to 4p\;^2S^o_{1/2}\to
4s\;^2P_{3/2}$. In the employed level scheme, a stepwise process is
dominant. That is, population transfer from metastable level $3d$
to upper laser level $4p$ takes place with subsequent emission on
the $4p \to 4s$ transition. Despite the parameters of metastable
level $3d\;^2P_{3/2}$ are not optimal (radiative lifetime of about
30~ns \cite{LUY,HIB} is comparable with lifetimes of $4p$-levels
involved in laser transitions), the conversion efficiency appeared
comparatively high ($>50\%$). In particular, it was much
higher than that in CW Raman lasers on atomic
gases~\cite{WEL2,WEL3}. In the present paper, we study factors that
can enhance output parameters of a Raman laser based on ions in a
gas discharge plasma. In the comparatively dense and cold argon
 laser plasma the key role is played by Coulomb ion-ion
collisions~\cite{BAB}.  Collisions with velocity change effectively
increase the number of particles that interact strongly with
resonant radiation, and the longer the level's lifetime the larger
increase in particle number there can be. Estimations indicate that
metastables with lifetimes $\ge 10^{-7}$~s should have width of the
hole ``burnt'' in the velocity distribution comparable to the
Doppler width. This means that nearly all ions in metastable state
interact with radiation.

In order to study collisional effects we achieved Raman
generation on a number of transitions in $\Lambda$-configuration
$3d \to 4p \to 4s(3d)$  starting from $3d'\,^2G, 3d\,^4F$ levels
which have almost no radiative decay \cite{LUY,HIB}. However, it is
 a fact that the metastable level relaxation in Ar$^+$-laser plasma
is determined by deactivation due to ion-electron collisions
\cite{BAB}.  Corresponding figures for levels of interest are
either unknown or there are contradictory data in different papers
\cite{WIL,ELB}. To carry out a quantitative analysis one has to
measure both excitation and deactivation rates of the metastable
level under real discharge condition. In this paper, we obtained
relatively high output parameters of Raman laser without  any
optimization (even a plain linear resonator scheme allowed the
up-conversion efficiency of more than 50\% with respect to pump
power, see section~\ref{CIL}).  Since the population transfer rate
and hence, the Raman laser output rely upon the
metastable level excitation rate and fraction of the velocity
distribution transferred to upper level, we concentrated at study
of absorbing transition for a specific $3d'\,^2G_{7/2}$ level,
aiming to clarify a role of collisions.  We studied the absorption
saturation (section~\ref{sut}) and the nonlinear resonance in the
probe field scheme brought about by the Bennett hole on the
metastable level (section~\ref{nelin}). Both experiments show a
giant broadening of the Bennett hole. Besides, the relaxation and
excitation rates for the metastable level were extracted from the
experimental data. We processed the acquired data using formulae
from~\cite{KURL}, which are valid for the simple two-level model. We
modified them considering two aspects:  difference between magnetic
sublevels and longitudinal inhomogeneity of the fields in otpically
dense media.  The formulae were checked against numerical simulation
(section~\ref{SISTEM}), which accounted for deceleration due to
dynamic friction force as well.

\section{Generation of the anti-Stokes Raman laser}\label{CIL}

In the present work, we have recorded the generation induced by
optical pumping from the metastable to the upper laser level on a
series of transitions:

\hspace{1cm} 610.4$\to$ 457.9 nm ($3d\;^2P_{1/2}\to
4p\;^2S^o_{1/2}\to 4s\;^2P_{1/2}$),*

\hspace{1cm} 648.3$\to$ 457.9 nm ($3d\;^2P_{3/2}\to
 4p\;^2S^o_{1/2}\to 4s\;^2P_{1/2}$),*

\hspace{1cm} 613.9$\to$ 496.5 nm ($3d\;^4F_{5/2}\to
4p\;^2D^o_{3/2}\to 4s\;^2P_{3/2}$),

\hspace{1cm} 624.3$\to$ 488.0 nm ($3d\;^4F_{7/2}\to
 4p\;^2D^o_{5/2}\to 4s\;^2P_{3/2}$),

\hspace{1cm} 617.2$\to$ 501.7 nm ($3d'\,^2G_{7/2}\to
4p'\,^2F^o_{5/2} \to 3d\;^2D_{3/2}$).

The schematic diagram of the experimental setup is presented in
fig.~1. We used the output of a single-frequency linearly polarized
 dye laser as the pump source. The pump radiation wavelength was
controlled by a $\lambda$-meter with 10$^{-3}$~nm resolution. This
was done using a fraction of the pump beam reflected by mirror
$M_1$. The main beam was directed by mirror $M_2$ and further
focused by lense $L$ into the discharge tube propagating along the
discharge through mirror $M_3$ of the argon laser which had a low
reflectance in the red spectrum range.  The cavity of the main
laser consisted of mirrors $M_3$ and $M_4$. In the absence of pump
beam, a usual series of well-known ArII lines was oserved.  Owing a
relatively large transmittance ($5\div 10\%$) of mirror $M_4$ in
the green-blue region, weak lines 457.9, 496.5 and 501.7~nm had low
gain-to-loss ratio. Under such conditions, the fraction of output
power that came from the optical pump was substantially larger than
the usual lasing. The output lines were separated by prism $Pr$.
Aperture spot $D_2$ served to select the TEM$_{00}$ generation mode
 of the argon laser and, together with $D_1$ maintained the pump
and Raman generation beams coaxiality, which simplified alignment a
lot.  In our experiments, we employed an argon ion laser with
longitudinal gas flow, in order to achieve a highly longitudinal
uniformity of the discharge~\cite{ODNOR}. 7~mm-bored discharge tube
was 1~m in length, it operated at current of 100~A, the pressure
being set optimally for argon laser generation. These parameters
were used throughout this work.

The scheme of generation on starred transitions is almost the same
 as in \cite{WEL}, i.e. same starting levels are used:
$3d\;^2P_{1/2}, 3d\;^2P_{3/2}$. Using instead of 437.5~nm line the
457.9~nm one that has a larger Einstein coefficient, we obtained
comparable to that of \cite{WEL} conversion efficiency (output
power exceeded $30\%$ of pump power), even in a plain linear scheme
of Raman laser. Farther, in the scheme with metastables~$^2G, ^4F$,
the efficiency was comparable to or higher than in \cite{WEL} (up
to~60\%).

For a detailed study, we chose the scheme with the largest pump
radiation absorption ($3d'\,^2G_{7/2}\to  4p'\,^2F^o_{5/2} \to
3d\;^2D_{3/2}$), see fig.~1. In the study of Raman generation on
 this transition, we carried out spectral measurements of the
501.7~nm line with the scanning interferometer $SI$ (5~GHz free
dispersion range) and photodiode $PD$ connected to an oscilloscope.
In the absence of optical pumping, we observed generation on
501.7~nm near the threshold, the line averaged gain coefficient
under these conditions amounted to about $\sim 10\%$, only a narrow
part of the gain contour about the line center was above the
threshold, a noise spectrum with the width of about~1~GHz being
recorded with random jumps among adjacent laser modes. Under a
single-frequency optical pumping on the 617.2~nm line an additional
narrow peak of Raman generation appeared. The noisy background
disappeared when the peak was tuned into the line center.
At the pump power of 70~mW we observed that the Raman peak shifts in
correspondence with detuning of the pump radiation and is present in
the output up to at least $\pm$4~GHz detuning. That is, even under
such detunings, the gain coefficient exceeded 10\% (induced by the
optical pump). This indicates a possibility to optimize the system
applying techniques of multi-mirror resonators as described in
\cite{WEL}, and produce a high output power having a broad tunable
range.

\section{Absorption saturation}\label{sut}

For absorption measurements the scheme of fig.~1 was used, but
mirror $M_4$ was removed, and hence, no generation occurred. Lens
$L$ was chosen so that it formed a beam only slightly diverging
along the discharge tube. The pump beam was attenuated with a set
of filters and the transmitted intensity was registered. The passed
through power was normalized to the average beam crossection $S =
(0.44 \pm 0.05)$~mm$^2$ and thus the average intensities $I_i$ and
$I_f$ at the entrance to and at the exit from the discharge tube,
correspondingly.

The saturation behavior of absorbed power for ionic transitions is
strongly dependent on diffusion in velocity space owing to Coulomb
collisions \cite{BAB,KURL}. Given the ion density $N_i = N_e \simeq
1.7\times 10^{14}$~cm$^{-3}$ and the temperature
 $T_i=m_iv_T^2/2\simeq 1$~eV, which correspond to the experimental
conditions, the effective frequency of ion-ion collisions amounts
to $\nu_{ii} \simeq 2\times 10^7$~c$^{-1}$\cite{BAB} and
corresponding diffusion coefficient is $D \simeq 5\times
10^{17}$~cm$^2$~c$^{-3}$

\begin{equation}\label{nuii} \nu_{ii} =
\frac{16\sqrt{\pi} N_i e^4 \Lambda} {3 m_i^2 v_T^3}, \hspace{1.5cm}
D = \frac{\nu_{ii}v_T^2}{2}
\end{equation}
where $\Lambda$ is the Coulomb logarithm, $e$ is the electron
 charge, $m_i$ is the ion mass.  It is shown in \cite{KURL} that at
strong diffusion the saturation of the absorbed power $P$ per unit
volume versus the field intensity $I = 16 \pi^2 \hbar c |G|^2 /
\lambda^3 A_{mn}, (|G|^2 = |E d_{mn}/2 \hbar|^2$ is Rabi frequency)
has a homogeneous form:

\begin{equation}\label{1}
P = {2 \hbar \omega \sqrt{\pi} (N_n(g_m/g_n)-N_m)\over
k v_T}{|G|^2\over 1+|G|^2/|G_s|^2},
\end{equation}

\begin{equation}\label{21}
|G_s|^2 = \frac{\sqrt{2 \nu_{ii}}k v_T}{2 \pi}
  \Biggl(  \frac{1}{\sqrt{\Gamma_m}} + \frac{1}{\sqrt{\Gamma_n}}
  \Biggr)^{-1}
\end{equation}
The formula is valid up to intensities determined by condition
$k v_T \sqrt{\nu_{ii}/2 \Gamma_n}  \gg |G|
\sqrt{\Gamma_{mn} /\Gamma_n}$, meaning that diffusional width
exceeds homogeneous one (including field broadening).
The results of \cite{KURL}
were obtained under assumption that

\begin{equation}\label{aproxim}
k v_T  \gg \Gamma_n, \Gamma_m, \Gamma_{mn}, \Omega;\hspace{1cm}
\nu_{ii} (k v_T)^2/ \Gamma_{mn}^3 \ll 1;
\end{equation}
where $\Omega = \omega -\omega_{mn}$ is the difference between the
radiation frequency and the Bohr frequency of the $m$--$n$
 transition, $\Gamma_j$ are relaxation constants for the levels,
$\Gamma_{mn} = (\Gamma_{m} + \Gamma_{n})/2$.
The latter condition holds well for the laser ArII lines. In our
case, this is not valid, since the metastable relaxation constants
$\Gamma_n$ are much less than those of the laser levels. Provided
the radiation is linearly polarized, we can neglect the degeneracy
and treat the $3d'\,^2G_{7/2} \to 4p'\,^2F^o_{5/2}$ transition as a
two-level system taking into consideration the degeneracy factors
$g_{m,n} = 2J_{m,n}+1$ of upper $(m)$ and lower $(n)$ levels.  To
check for validity of expression (\ref{1}), (\ref{21}) for concrete
levels we compared them with numerical solution of the equations
for the density matrix (see Sec.~\ref{SISTEM}).

As long as up to 90\% of the incident radiation is absorbed, it is
necessary to consider the optical density of the medium. In such
 case the relation between the incident ($I_i$) and output ($I_f$)
intensities has the form:

\begin{equation}\label{3}
    \ln \biggl( \frac{I_i}{I_f} \biggr) + \frac{I_i - I_f}{I_s} =
    k_o l,
\end{equation}
where the small-signal absorption coefficient has usual form

\begin{equation}\label{31}
k_o \simeq  \frac{\lambda^3 A_{mn} }
{8 \pi^{3/2} v_T} \Bigl( N_n(g_m/g_n) - N_m  \Bigr),
\end{equation}
and the saturation intensity is

\begin{equation}\label{32}
I_s = \frac{8 \pi \sqrt{2 \nu_{ii}} k v_T \hbar c}
      {\lambda^3 A_{mn}}
     \Biggl(  \frac{1}{\sqrt{\Gamma_m}} +
     \frac{1}{\sqrt{\Gamma_n}} \Biggr)^{-1}.
\end{equation}

In the fig.~2, the dependence of the absorption $\Delta I=I_i -
 I_f$ on the incident intensity $I_i$ given by the expression
(\ref{3}) is plotted along with the experimental points according
to maximum-likelihood fit. The extracted values of the unsaturated
absorption coefficient $k_o = (2.14 \pm 0.05) \times 10^{-2}$
cm$^{-1},$ and saturation intensity $I_s = 11.2 \pm 2.1$
${\mbox{W}}/{\mbox{cm}^2}$.
A comparison between the absorption on this
transition and the gain coefficient for the adjacent transition
501.7~nm indicated that the metastable level $3d'^2G_{7/2}$
 population is 14 times the $4p'^2F^0_{5/2}$ level population.
Knowing the Einstein coefficient for this transition
$A_{mn} = 2.2\times 10^7$~c$^{-1}$ \cite{HIB}
and the discharge length being $l=100$cm,
one can obtain the
lower level population $N_n \simeq (5.33 \pm 0.14)\times
10^{10}$~cm$^{-3}$ from formula (\ref{31}).
One can also determine the lower level
decay rate with the help of Eq.(\ref{32}). The upper level decay
 rate was taken from literature: $\Gamma_m = 2.0 \times
10^8$~c$^{-1}$ ($\tau_m = 8.4$~ns -- radiative lifetime \cite{HIB},
$K_m \simeq 10^{-7}$$\mbox{cm}^3\mbox{c}^{-1}$ is the deactivation
constant \cite{JOL}). The obtained lower level lifetime is
$\Gamma_n \simeq (7.7 \pm 4.6) \times 10^7$c$^{-1}$, the low
accuracy is owing to the fact that only slight saturation was
achieved. Additional error is introduced by uncertainty in the
measured saturation intensity caused by transverse nonuniformity of
the beam. Since the expression for $\Gamma_n$ includes a difference
of two very close values, the calculated result is rather an
estimation. Measuring the width of the nonlinear resonance in the
spectrum of the probe field one can determine more accurately the
metastable level relaxation constant. As soon as the contribution
of the levels $m$ and $n$ into the nonlinear resonance is inversely
proportional to $\Gamma_j$, one can draw a conclusion, on the basis
of the yielded results, that the contribution of the lower level is
at least 3 times greater that of the upper one.  The measurements
of the nonlinear resonance shapes are covered in the next section.

\section{Nonlinear resonance}\label{nelin}

The scheme of the experiment on nonlinear resonance measurements in
 the probe field spectrum is presented in fig.~3. To split the dye
laser beam into two, the pump beam and the probe one (a and b in
fig.~3, correspondingly), we used a reflection from the uncoated
surface of the output mirror ($M_1$ in fig.~3). The probe beam
power amounted to a few percent of the strong one. Lenses $L_1$ and
$L_2$ were used to ensure minimal divergence of the beams along the
discharge tube, the diameter of the probe beam being 3 times
smaller than that of the pump beam.  The passed through the
discharge probe beam was detected behind the beam splitter $BS_2$
(transmittance $t \sim 20 \%$). Lock-in detector Unipan~232B was
used. A fraction of the pump beam coming through the
splitter~$BS_1$ and modulated by chopper~$Ch$ was used as the
reference signal.  Chopper in position~1 modulates both beams, and
the lock-in detector registers the transmitted through the
discharge power of the probe beam. Placed in position~2, the
chopper offects the pump beam only, and the lock-in detector
registers variation of the passed through probe power occurring due
to influence of the pump beam. The intensity of the pump field was
measured with an account for change in the dye laser beam
crossection. To do this an additional registration of variation in
intensity near the beam axis was made using  an aperture stop $D$
(the diameter is much smaller than the effective beam diameter).
The frequency detuning was measured with scanning
interferometer~$SI$, photodiode~$PD$ and oscilloscope~$Os$.
Along with the dye laser frequency the scanning interferometer
registered the single-frequency He-Ne laser signal. The latter
served as a reference point to detuning of the dye laser.

Owing to a strong absorption, the depth of the ``burnt'' Bennett
 hole changes along the discharge. That's why we need a formula for
the resonance in an optically dense medium. Neglecting the
influence of the probe field $I_1$ on the strong one $I_2$ which
propagates towards $I_1$ in the opposite direction from the point
$z=l$ to $z=0$, we have an exponential intensity dependence on the
linear distance $z$:  $I_{2}(z) = I_{2}(l)\exp{[k(\Omega_2)(z -
l)]}$. The corresponding dependence of the weak field intensity
$I_1(z)$ can be extracted from the following equation, taking into
account a weak saturation of the transition by the strong field
$I_2(z)$:

\begin{equation}\label{4}
    \frac{dI_{1}(z)}{dz} = - k(\Omega_1) \cdot I_{1}(z)
    \Bigl( 1 - F( \Omega_1, \Omega_2 )\frac{I_{2}(z)}{I_s} \Bigr),
\end{equation}
where $\Omega_{1,2}$ are frequency detuning of the weak and strong
fields from the exact resonance correspondingly;

\begin{equation}\label{5}
F( \Omega_1, \Omega_2 ) \simeq \exp{ \Bigl(-
\frac{|\Omega_2-\Omega_1|}{k v_T} \sqrt{\frac{2
\Gamma_n}{\nu_{ii}}} \Bigr)}
\end{equation}
where $F$ is the function that describes the shape of the Bennett
 hole in case of diffusion broadening (see review \cite{BAB} and
references therein, validity of the expression as applied to the
studied transition is discussed in Section~\ref{SISTEM}). In the
investigated scheme $\Omega_1 = - \Omega_2 = \Omega$, thus
velocities of particles resonant to probe and pump field $ \pm
\Omega /k$ only differ in sign).

The solution of Eq.(\ref{4}) can be written in the following form:

\begin{equation}\label{10}
    F_{1,2} = F( \Omega, -\Omega )
    \simeq \frac{I_s (I_{1}(l) - \mbox{\~I}_{1}(l))}
     {I_{2}(l) \mbox{\~I}_{1}(0) ( 1- \exp{( - k(\Omega) l )} )
        \exp{( - k(\Omega) l )}},
\end{equation}
where $\mbox{\~I}_{1}(l) = \mbox{\~I}_{1}(l) \exp{( - k(\Omega)l
 )}$ is the probe field intensity in the absence of the strong
field, i.e. at $I_2=0$.

In the experiments, the readings of the lock-in detector were
recorded when the chopper $Ch$ was in position~2; in this case they
were proportional to $I_{1}(l) - \mbox{\~I}_{1}(l)$. The input
intensities of the probe $\mbox{\~I}_{1}(0)$ and pump $I_{2}(l)$
 fields were proportional to the output power of the dye laser.
With the chopper in position~1 we measured the dependence of the
absorption coefficient upon detuning $k(\Omega) = k_o
\exp{(-\Omega^2/(k v_T)^2)}$.  From these data we reconstructed
$F_{1,2}$ as a function of detuning $\Omega$ up to an arbitrary
amplitude.
The dependence calculated in this way is demonstrated in fig.~4.
It's relevant to mention that the experimental data consist of two
overlapping sets of points, because it was difficult to cover the
required range in one sweep. Experimental points were approximated
by expression (\ref{5}) using the maximum-likelihood fit. Three
parameters were fitted: magnitude, width and detuning zero shift.

The processed data yielded diffusion full width at half a maximum
(FWHM) $\Delta = (\ln{2}/ \pi) \sqrt{\nu_{ii}/2 \Gamma_n} k v_T =(
 3.0 \pm 0.3)$~GHz, the Doppler width (FWHM) being $\Delta \nu_D =
k v_T \sqrt{\ln {2}}/ \pi =(5.3 \pm 0.1)$~GHz. Internal noises of
the dye laser in the vicinity of the strong field chopping
frequency can lead to systematic error, that is, the total error
can exceed 10\%.  The decay rate of the metastable level extracted
from the diffusion width is $\Gamma_n = (2.3 \pm 0.5) \times
10^{-7}$c$^{-1}$, which agrees with the value obtained from
saturation intensity. Hence, our assumptions concerning
contributions of the upper and lower levels into the resulting
contour of the nonlinear resonance are obviously true,
$\Gamma_n\simeq 0.1\Gamma_m$. To compare the Bennett hole width and
that of the Doppler contour, we showed in the inset to fig.~4 the
corresponding velocity distributions at zero detuning from exact
resonance. We should take into account that the Bennett hole width
is $\Delta$ and the width of function $F_{1,2}$ in Eq.(\ref{5}) is
$\Delta /2$ which comes from scanning the hole with double speed.
It is seen from the plot that collisions ``blur'' the Bennett hole
to almost fill the whole Doppler contour.

Besides, knowing the relaxation rate and population we find that the
excitation rate of the metastable level $Q_n$  $1.6$ times greater
 than that of the upper laser level.

\section{Numerical simulation}\label{SISTEM}

As soon as not all the assumptions made in \cite{KURL} hold in
case of our investigations, we had check for accuracy of the
 analytical expressions. To do this we carried out a numerical
solution of the equation array for  the density matrix of a
two-level system in the field of a running electromagnetic wave
including Coulomb diffusion and dynamic friction force. The array
to solve follows \cite{BAB}:

\begin{eqnarray}
(\Gamma_{mn} -i\Omega +i k v)\rho_{mn} - \hat{D}\rho_{mn} =
  -i G(\rho_{mm} - \rho_{nn}), \nonumber\\
\Gamma_m\rho_{mm} - \hat{D}\rho_{mm} = Q_mW(v) -
  2\mbox{{\bf Re}}(i G^* \rho_{mn}), \label{SIST}\\
\Gamma_n\rho_{nn} - \hat{D}\rho_{nn} = Q_nW(v) +
  2\mbox{{\bf Re}}(i G^* \rho_{mn}); \nonumber
\end{eqnarray}
where
\begin{equation}
\hat{D}\rho_{ij} =
\nu_{ii} \frac{\partial}{\partial v}(v\rho_{ij}) +
\frac{1}{2}\nu_{ii} v_{\mbox{\tiny T}}^{2}
 \frac{\partial^2}{\partial v^2}\rho_{ij}, \end{equation} $Q_j$ is
total excitation rate of $j$-th level, $W(v)$ is one-dimensional
Maxwell distribution over velocity having width $v_{\mbox{\tiny
T}}$. We neglected terms proportional to $A_{mn}$, since they are
small ($A_{mn} \simeq 0.1 \Gamma_m$) for the studied transition.
For the numerical solution of Eq.(\ref{SIST}) at preset parameters
in operator $\hat{D}$ symmetrical differences were substituted for
partial velocity derivatives. At $v = 5 v_{\mbox{\tiny T}}$ we took
the asymptotic condition $\rho = 0,$ and solved the resulting array
of linear algebraic equations using the matrix sweep method.  To
calculate the field work and the shape of the nonlinear resonance,
we averaged over velocities the produced distribution for
$\rho_{ij}(v)$ using formalism given in \cite{BAB}.

Different behavior of the magnetic sublevels was taken
into consideration, too.  When a linearly polarized field interacts
 with transition $3d'\,^2G_{7/2}\to 4p'\,^2F^o_{5/2}$ only allowed
transitions are those preserving the projection of magnetic
momentum $M$ onto the polarization direction.  Magnetic sublevels
with the same projection $M$ can be regarded as independent
two-level subsystems with different dipole moments:

\begin{equation}
       |d_M|^2 = |\langle m||d||n\rangle |^2
                                 \left| \begin{array}{c}
                                 \left( \begin{array}{ccc}
                                   7/2  &  5/2  &  1   \\
                                   M    &  -M   &  0
                                   \end{array} \right)
                                   \end{array} \right|^2
\end{equation}
where $\langle m||d||n\rangle $ is the normalized dipole moment,
\begin{equation}\label{dipol}
  |d_{5/2}|^2 = \frac{3}{28}\frac{|\langle m||d||n\rangle |^2}{3};
  |d_{3/2}|^2 = \frac{5}{28}\frac{|\langle m||d||n\rangle |^2}{3};
  |d_{1/2}|^2 = \frac{3}{14}\frac{|\langle m||d||n\rangle |^2}{3}
\end{equation}
At a given intensity $I$, the field magnitude (in frequency units)
$|G_M|^2 = |E d_{M}/2 \hbar|^2$ is different for every subsystem.
 The absorbed power is found by summing up over all subsystems. For
each of them expression (\ref{1}) is valid with the substitution
$|G|^2 \to |G_M|^2$ and $[(g_m/g_n)N_n-N_m] \to
[(N_n/g_n)-(N_m/g_m)]$. At low and high intensities $I \ll I_s$ and
$I \gg I_s$, the total absorption is exactly the same as that of a
two-level system according to Eq.(\ref{1}). The largest deviation
is achieved at $I = I_s$ and does not exceed to 2\%.

Numerical calculations were carried out  for values for the
 population of the metastable level acquired in experiments on
absorption saturation, and values for the upper level population
from measurements of the gain coefficient of a laser line on the
adjacent transition. The metastable level decay rate was taken from
experiment with nonlinear resonance in the probe field scheme.

In fig.~5a, a comparison between the dependence of absorbed power
 $P$ per unit volume upon field intensity $I_i$ for a two-level
system in an optically thin medium according to Eq.(\ref{1}) (the
dotted line) and that calculated on a PC with an account for
dynamic friction and specific values for dipole moments of magnetic
sublevels (\ref{dipol}) (the solid curve).  As can be seen from
fig.~5a, the difference between the two curves is within 5\%. This
means that dynamic friction is the main source of discrepancy.
Besides, the numerical data at given values $\Gamma_j, Q_j, A_{mn},
\nu_{ii}, v_T$ can be fitted by the analytical expression $P =
k_o I/(1+I/I_s),$ but with different values for $k_o, I_s.$ The
yielded fit parameters are lower by 2\% and 6\% correspondingly
than $k_o,$ $I_s$ in (\ref{31}) and (\ref{32}).

The contour of the nonlinear resonance $\Delta k(\Omega)$ was
calculated numerically for intensity of the strong field $I_2 = 0.1
I_s,$ and probe field intensity $I_1 \ll I_s$ in accordance with the
experimental conditions. In fig.~5b, the solid curve is for
numerical calculations and the dashed line is for analytical
approximation $\Delta k(\Omega) = F_{1,2} \exp{(-\Omega^2/(k
v_T)^2)} {I_2}/{I_s}$ used for  experimental data processing. It
deviates from the numerical simulation by a factor of 1.45 with an
accuracy not worse than 3\% over all the detuning range, i.e.
copies the resonance shape.

\section{Discussion}

We analyzed the dependence of absorption upon pump field intensity,
 as well as the nonlinear resonance that comes from the Bennett
hole on the metastable level. This included approximation of the
experimental values with analytical expression and a comparison of
the theory with the numerical solutions of the equation array for
the density matrix.  As a result, it was clearly demonstrated that
both the shape of the saturation curve and shape of the Bennett
hole are primarily a consequence of Coulomb diffusion of ions in
the velocity space. The diffusion leads to an abnormal broadening
of the Bennett hole:  diffusional width of the hole $\Delta$ for
the metastable $3d'\,^2G_{7/2}$ reaches 3~GHz, whereas Lorentzian
width of the transition $\Gamma_{mn}/\pi$
amounts to as little as 35~MHz. This means that is a relative
broadening by a factor of nearly 100.  With the field broadening
included, the transition width $\Gamma_{mn} \sqrt{1+ 4
|G|^2/\Gamma_m \Gamma_n} \sim 10 \Gamma_{mn}$ is also much less than
the diffusional one. Thus, the giant Coulomb broadening completely
determines the character of saturation on the transition.  Instead
 of the well-known ``hole burning'' in the velocity distribution
one observes nearly complete saturation of the whole Doppler
contour.  The form of the saturation resonance has an unusual
exponential cusp feature (see fig.~4). It's relevant to note that
such exponential wings were observed in the spectrum of the probe
field on transition $3d'\,^2G_{9/2} \to 4p'\,^2F_{7/2}$ of argon
ion under conditions of hollow cathode discharge \cite{ELB,ELB2}.
The authors of these papers attributed the broadening to ion-atom
collisions.  Unfortunately, plasma parameters were not measured and
this is a serious obstacle to quantitative analysis. However,
despite considerable difference in plasma parameters (specifically,
much lower charged particle concentrations in hollow cathode),
Coulomb collisions may also play a noticeable role there.

Coulomb broadening changes completely the form of the saturation
 curve.  The typical for single-frequency pumping inhomogeneous
saturation $k_o\Big /\sqrt{1+|G|^2/|G_o|^2}$ gets substituted by
homogeneous saturation: $k_o\Big /(1+|G|^2/|G_s|^2)$. The
saturation intensity differs by a factor of 60 for these cases. The
dotted lines in fig.~2 correspond to inhomogeneous saturation and
to the asymptote of the maximum absorbed power $P \simeq (6/8) Q_n
\hbar \omega S l$ (the expression holds when $\Gamma_n \ll
\Gamma_m$). At intensities used in our experiments, the curve that
takes Coulomb diffusion into account saturates noticeably nearing
the asymptotic value, whereas the inhomogeneous saturation curve is
far from this.  At pump intensities $I\sim I_s$ diffusion results
in absorption increase (and hence, the output power of the Raman
laser on the adjacent transition) by a factor 3--5 depending on
specific values of relaxation constants. Note, that the
perturbation influence of Raman generation on absorption of the
pump radiation is small in case $\Gamma_n \ll \Gamma_m$. At low
intensities (much lower than the saturation intensity on the
adjacent transition) the generation repeats the dependence of
absorbed power on the pump input.  We should mention that earlier
\cite{RAUT}, simple formulae for the output power of Raman laser
using perturbation method in the pump and generation fields were
already written that qualitatively demonstrate the role played by
Coulomb diffusion.

Let's compare the obtained value for the relaxation constant of the
metastable level $\Gamma_n \simeq 2 \times 10^7$~c$^{-1}$ with data
found in other papers. We know no published figures for level
$3d'^2G_{7/2}$ deactivation in plasma, however there are some
 results on the similar metastable level $3d'^2G_{9/2}$. In
\cite{WIL}, the laser induced fluorescence was observed in a hollow
cathode discharge.  The measured electron deactivation coefficient
appeared to be $(2.5 \pm 1.0) \times 10^{-8}$~cm$^3$/c, electron
temperature being  $T_e = 3.5$~eV. For electron density $N_e \simeq
2 \times 10^{14}$~cm$^{-3}$, which corresponds to our experiments,
we get the decay rate $\Gamma_n \sim 5 \times 10^6$~c$^{-1}$. In
\cite{ELB}, there is the relaxation constant of the metastable
level $3d'^2G_{9/2}$ $\Gamma_n \sim 10^8$~c$^{-1}$, again for a
hollow cathode discharge. Authors of Ref.  \cite{ELB} supposed that
metastable is quenched due to resonant charge-exchange collisions
of excited argon ions with atoms in ground state.  Estimations for
electron deactivation of level $3d'^2G_{7/2}$ that take into
consideration only the predominant channel $3d'^2G_{7/2} \to
4p'\,^2F^o_{5/2}$ done in Bates-Damgaard approximation (using
tables in the book by Vainshtein {et. al.} \cite{VSJU}) yields
$\Gamma_n \simeq 4 \times 10^6$c$^{-1}$, which is slightly lower
than the measured value.  Probably, in our conditions relaxation
occurred due to both collisions with electron and resonant
charge-exchange with neutrals.  This shows that the experimental
value for the relaxation rate $\Gamma_n \simeq 2 \times
10^7$c$^{-1}$ of level $3d'^2G_{7/2}$ does not contradict to
estimations and lies within scatter of data in \cite{WIL,ELB} for
the relaxation rate of the similar level $3d'\,^2G_{9/2}$.

Note, that the studied metastable level has the relaxation rate
comparable with ion-ion collision frequency. Under such
conditions the dynamic friction force and velocity dependence of
the diffusion coefficient can lead to a noticiable narrowing
of the Bennett hole centered at a wing of velocity distribution
$v > v_T$ \cite{Step}. Experiments with independently tuned
strong and probe fields may test the phenomenon.

\section{Conclusion}

1. The present work has demonstrated the possibility of Raman
generation due to an optical pumping from
metastable levels $3d'\,^2G, 3d\,^4F$ on a number of generation
lines of argon laser.

2. For absorbing transition $3d'\,^2G_{7/2}\to 4p'\,^2F^o_{5/2}$ of
the Raman laser $617.2\to 501.7$~nm, the unsaturated absorption
coefficient $k_o = 2 \times 10^{-2}$ cm$^{-1},$ and saturation
intensity $I_s = 11$ ${\mbox{W}}/{\mbox{cm}^2}$ have been
determined.

3. The ratios of populations ($\sim 14$) and excitation rates
($\sim 1.6$) for levels $3d'\,^2G_{7/2}$ and $4p'\,^2F^o_{5/2}$ have
been measured. The relaxation rate of level $3d'\,^2G_{7/2}$
$\Gamma_n = 2.3 \times 10^{-7}$c$^{-1}$ is determined primarily by
collisional deactivation in plasma.

4. The Bennett hole on metastable level $3d'\,^2G_{7/2}$ is
broadened by a factor of 100 owing to Coulomb collisions. The hole
has exponential shape and its width $\Delta \sim$3~GHz is
comparable to the Doppler one. This giant broadening is the key
factor in the character of saturation on the transition. Instead of
usual ``hole burning'' in the velocity distribution, we observed an
almost complete saturation of the whole Doppler contour. This adds
up to the population transfer to the upper level and hence, to the
efficiency of the Raman laser.

\section*{Acknowledgements}

The authors are grateful to S.M.~Kobtsev for technical assistance
 and to E.A.~Yukov for fruitful discussions.

The present paper has been partially sponsored by the International
Science Foundation and Russian Government, grant RCN~300.

\newpage
\section*{List of captions}

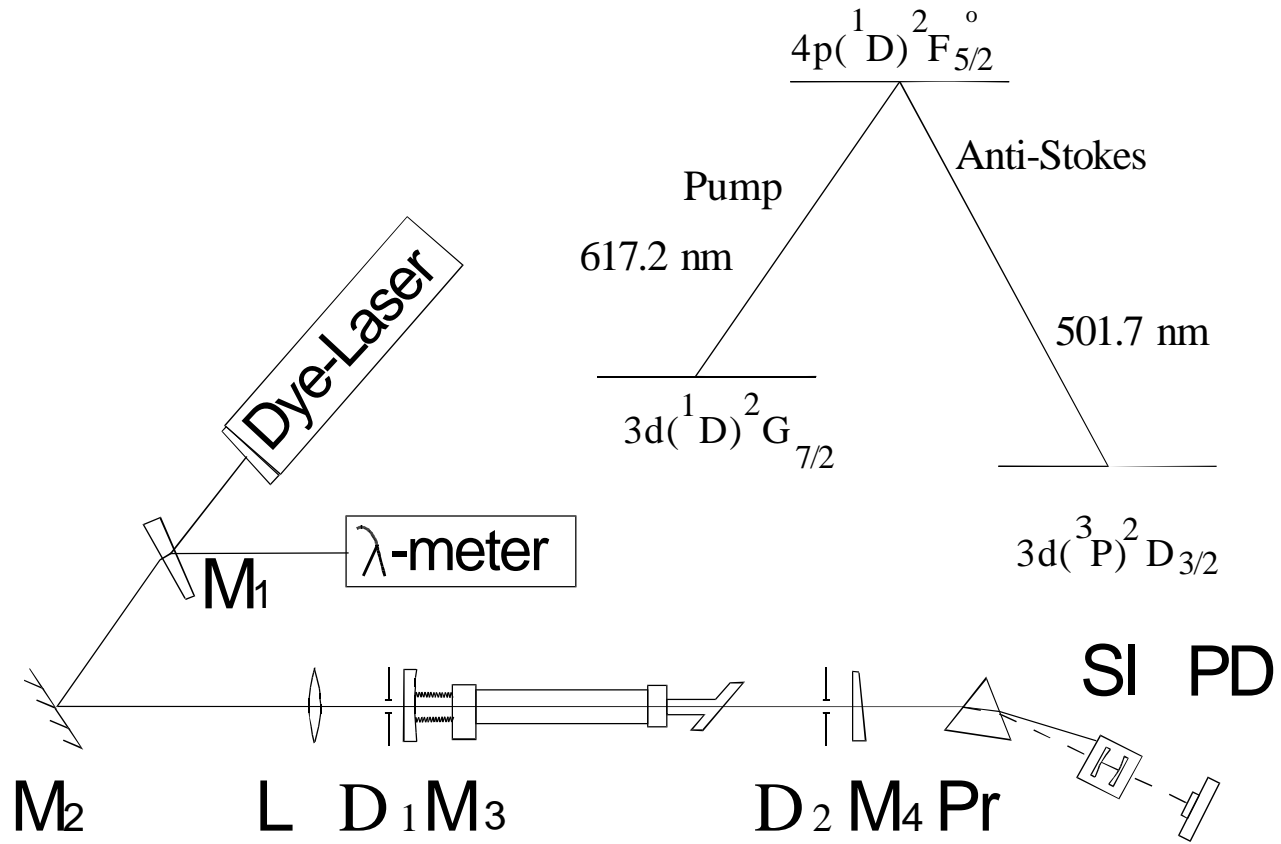
Fig.~1. Experimental setup for the study of Raman generation:
$M_1$ -- low-reflectance mirror;
$M_2$ -- high-reflectance turning mirror; $\lambda$-meter --
wavelength controller, $L_1$ -- lens;
$D_1, D_2$ -- aperture stops;
$M_3, M_4$ -- argon laser cavity mirrors;
$Pr$ -- dispersion prism;
$SI$ -- scanning interferometer;
$PD$ -- photo detector. There is a scheme of levels involved in
Raman generation in this figure.

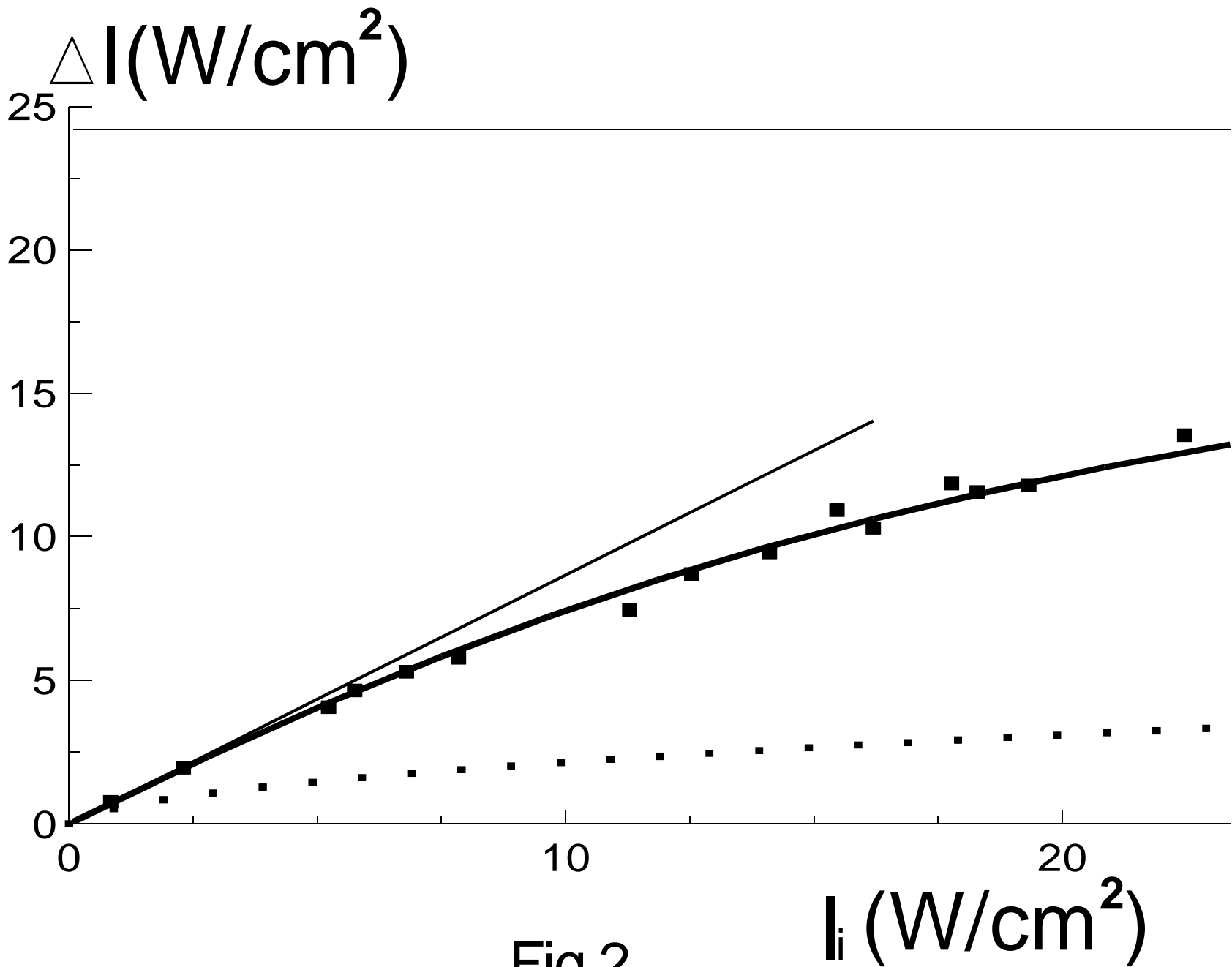
Fig.~2. Dependence of the absorbed intensity $\Delta I =I_i-I_f$ on
the transition $3d'\,^2G_{7/2}\to 4p'\,^2F^o_{5/2}$ upon the
incident intensity $I_i$. Boxes correspond to experimental data.
Theoretical dependence (solid curve)~(\ref{3}) is fitted
with the maximum-likelihood algorithm. Dots correspond to
inhomogeneous saturation with Coulomb diffusion neglected.
The strait lines correspond to the derivative from the saturation
curves at zero intensities, and to the asymptote of maximum possible
absorption. (Discharge length~100~cm, electron density
$N_e \simeq 1.7 \times 10^{-14}$cm$^{-3}$, electron
and ion temperatures $T_e \simeq 4$eV, $T_i \simeq 0.8$eV,
respectively).

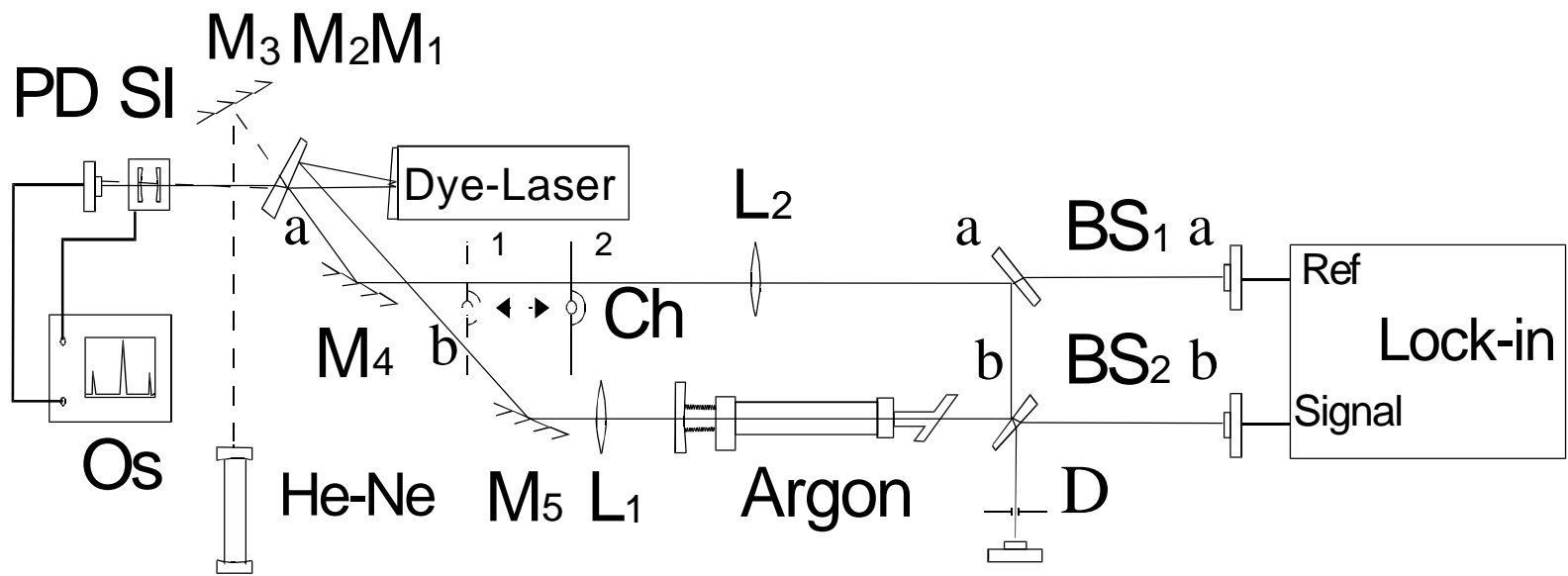
Fig.~3. Experimental setup for study of nonlinear resonance in the
probe field configuration:
$M_1$ -- coupling mirror of dye laser,
$M_2, M_3, M_4, M_5$ -- beam turning mirrors,
$BS_1, BS_2$ -- beam splitters, $L_1, L_2$ -- focusing lenses,
Lock-in -- lock-in detector, $Ch$ -- beam chopper, D -- small
aperture stop.
Letters {\em a} and {\em b} mark pump and probe beam paths
correspondingly. Chopper in position~1 modulates both pump and probe
fields, in position~2 it only effects the pump beam.
The scanning interferometer ($SI$), photo detector
($PD$), oscilloscope ($OS$) and single-frequency He-Ne~laser are the
components of the spectrum analyzer.

Fig.~4. Function (\ref{10}) $F_{1,2}$ versus detuning $\Omega
/(2\pi)$ (in GHz)  fitted into the experimental points by
maximum-likelihood method.  Crosses and boxes are for two different
data sets. The discharge parameters are the same as in the
experiment on saturation measurement.  For comparison, the Doppler
velocity distribution (in $v_T$ units) and the shape of the Bennett
hole on the metastable level are also shown.

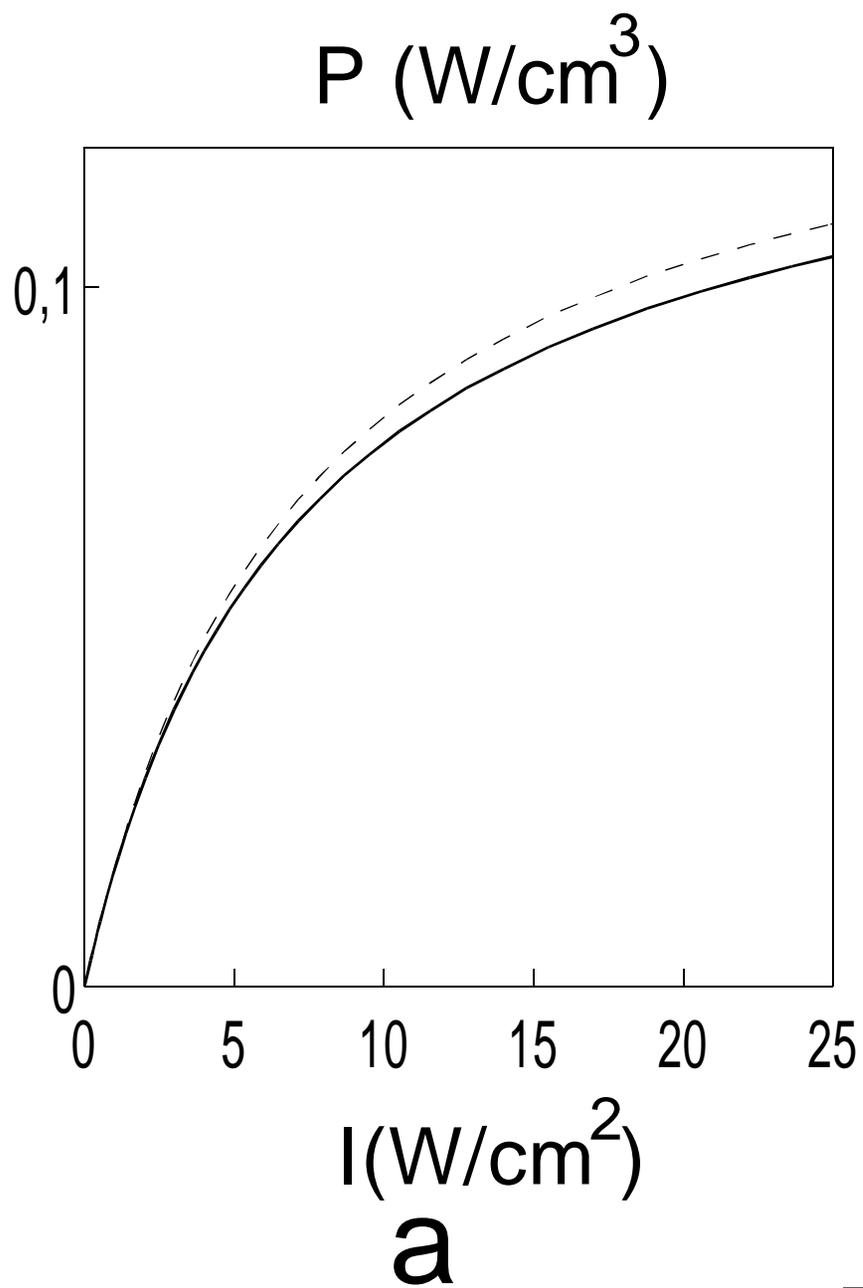
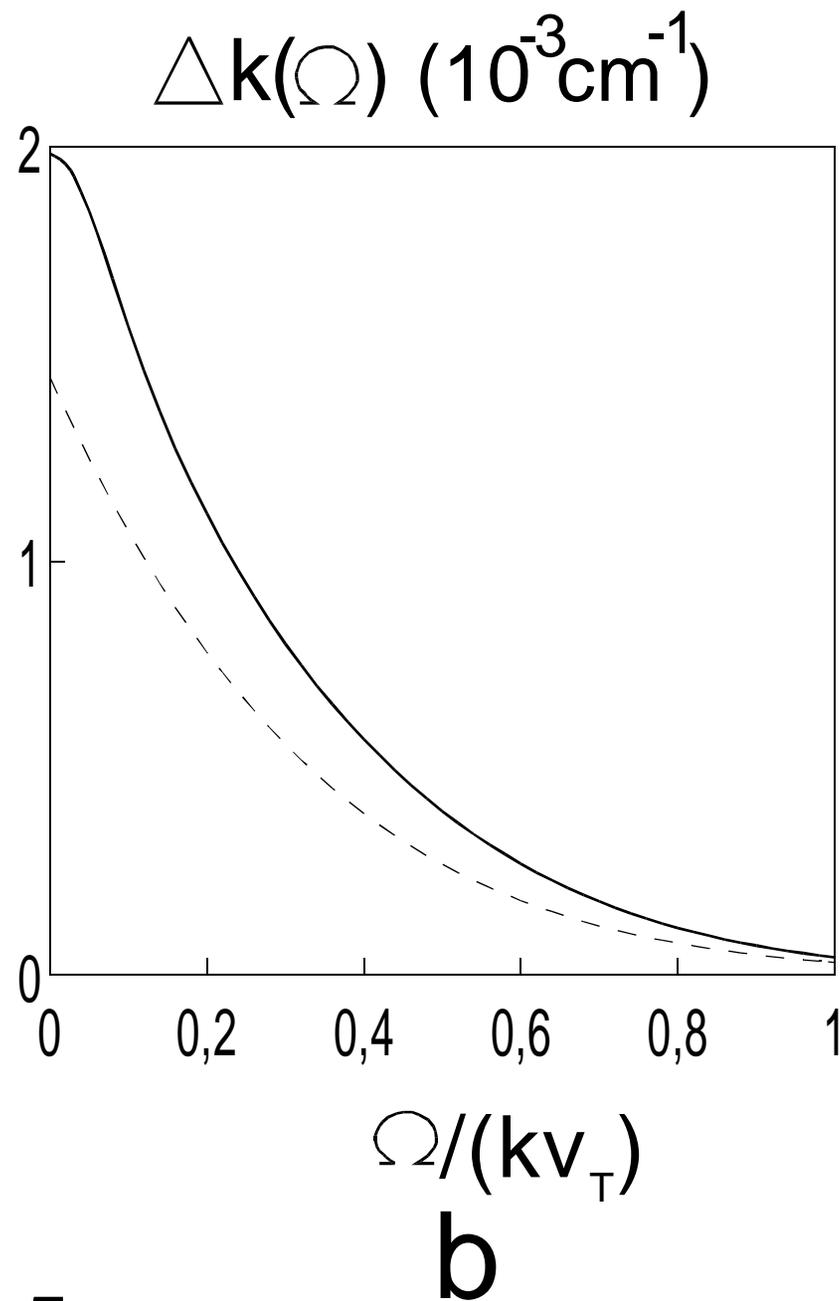
Fig.~5. a  -- Dependence of the absorbed power P upon field intensity
$I_i$ in an optically thin medium for a two-level system (\ref{1})
and a numerical simulation (\ref{SIST}) for magnetic sublevels with
friction force included (dashed and solid curves correspondingly).
b -- The variation of the absorption coefficient $\Delta k(\Omega) =
F_{1,2} \exp{(-\Omega^2/(k v_T)^2)}{I_2}/{I_s}$ caused by the strong
field at the intensity $I_2 = 0.1 I_s$: solid curve plots the
results of numerical simulation, dashed curve is the analytical
approximation (parameters taken from experiments).

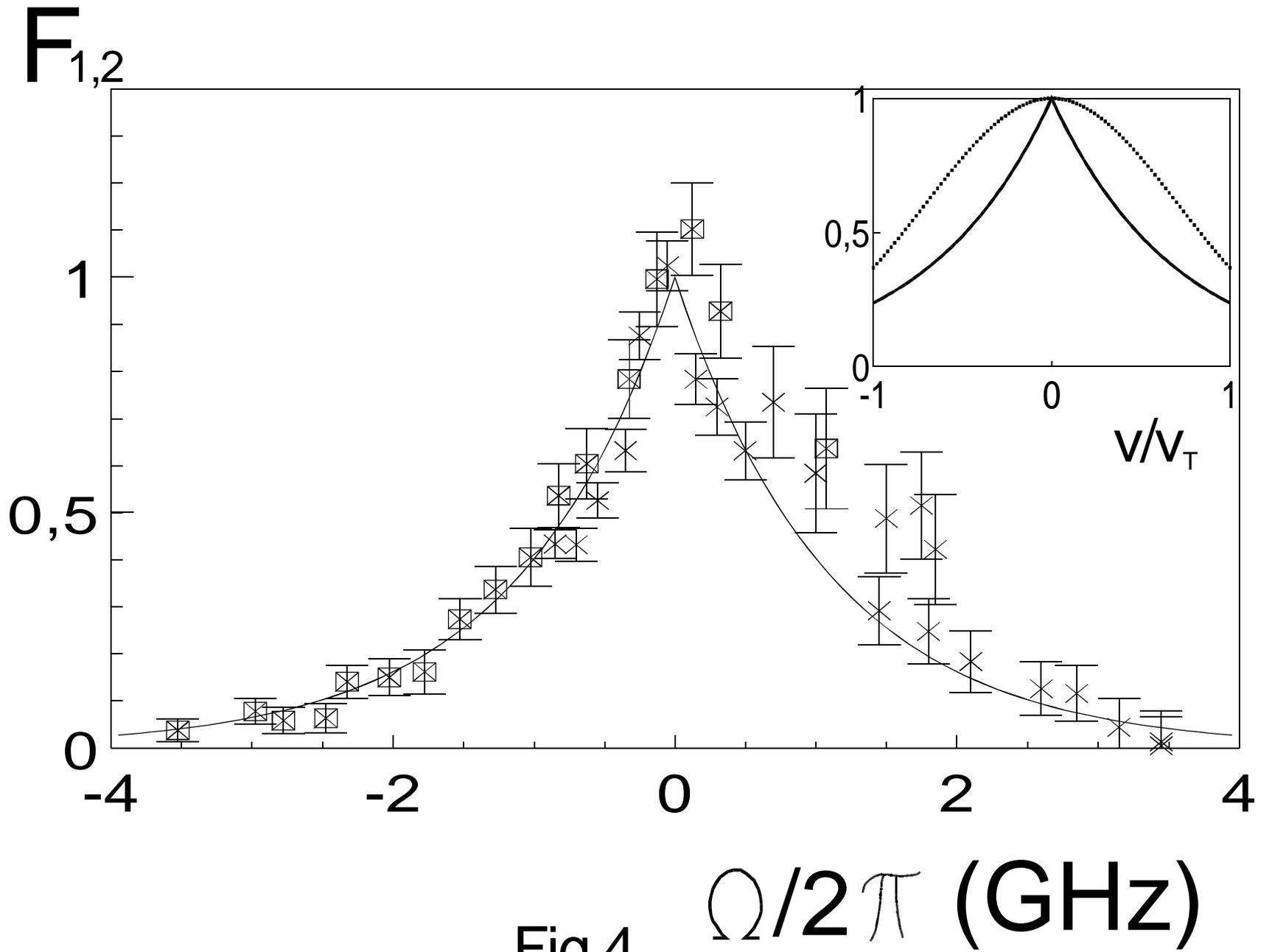
\end{document}